\pdfoutput=1
\documentclass[%
showpacs,
 amsmath,amssymb,
 aps,
superscriptaddress,
prl,
floatfix,
twocolumn, %
]{revtex4-1}

\usepackage{graphicx}
\usepackage{dcolumn}
\usepackage{bm}
\usepackage[utf8]{inputenc}
\usepackage{amsmath}
\usepackage[amssymb]{SIunits}
\usepackage{amsfonts}
\usepackage{amssymb}
\usepackage{mathtools} 
\usepackage{dsfont}
\usepackage{color}
\usepackage{isotope}
\usepackage{calc}
\usepackage{braket}

\usepackage{mathrsfs} %

\begin{document}


\title{Entropy of the BEC Ground State: Correlation vs Ground State Entropy} %
\author{Moochan B.~Kim}
\affiliation{Texas A\&M University, College Station, TX 77843} %
\author{Anatoly Svidzinsky}
\affiliation{Texas A\&M University, College Station, TX 77843} %

\author{Girish S.~Agarwal}
\affiliation{Texas A\&M University, College Station, TX 77843} %

\author{Marlan O.~Scully} %
\affiliation{Texas A\&M University, College Station, TX 77843} %
\affiliation{Baylor University, Waco, TX 76706} %

\date{\today}

\begin{abstract} %
Calculation of the entropy of an ideal Bose Einstein Condensate (BEC) in a three dimensional trap %
reveals unusual, previously unrecognized, features of the Canonical Ensemble. %
It is found that, for any temperature, the entropy of the Bose gas is equal to the entropy of the excited particles %
although the entropy of the particles in the ground state is nonzero. %
We explain this by considering the correlations between the ground state particles and particles in the excited states. %
These correlations lead to a correlation entropy which is exactly equal to the contribution from the ground state. %
The correlations themselves arise from the fact that we have a fixed number of particles obeying quantum statistics. %
We present results for correlation functions between the ground and excited states in Bose gas, %
so to clarify the role of fluctuations in the system. %
We also report the sub-Poissonian nature of the ground state fluctuations. %
\end{abstract} %


\maketitle

\textit{Introduction}---The properties of a Bose condensate \cite{Ref:PethickSmithBEC, Ref:Ziff} %
are usually studied using a grand canonical ensemble by making a number of assumptions which can be justified in the thermodynamic limit \cite{Ref:Huang2nd, Ref:Pathria3rd, Ref:LandauStatPhy}. %
For a condensate consisting of relatively small number of particles, %
it is better to use a canonical ensemble. %
This ensemble is useful in understanding the particle number distribution, %
as well as the fluctuations in the number of particles in ground states and excited states, has been obtained %
\cite{Ref:WeissWilens, Ref:NavezPRL, Ref:IdziaszekPRL, Ref:KocharovskyPRL, Ref:HolthausAnnPhys}. %
Such calculations do not require thermodynamic limit. %
An important result is the distribution of the number of particles in the ground state. %
Recent work presents the entropy of the ground state of an ideal $N$ particle Bose-Einstein condensate (BEC) %
from the condensate density matrix \cite{Ref:CNBseries, Ref:ScullyLaserQuantumEntropy} %
\begin{align} %
\label{Eq:DMBECGnd} %
\rho_{n_0n_0} = \frac{ \mathcal{H}^{N-n_0}}{(N-n_0)!} e^{-\mathcal{H}}, %
\end{align} %
where $\mathcal{H} = N (T / T_c)^3$ for a harmonic trap at temperature $T$ %
and critical temperature $T_c$ and $n_0$ is the number of atoms in the condensate state. %

This distribution has some novel features--it is like the well known laser distribution for photons in a single mode laser. %
This distribution can be used to calculate the thermodynamic properties of the ground state; %
in particular the approximate expression for entropy was obtained.
From the von Neumann entropy %
\begin{align} %
\label{Eq:vonNeumannEntropy} %
S = - k_\text{B} \sum_n \rho_{nn} \ln \rho_{nn}, %
\end{align} %
with Boltzmann constant $k_\text{B}$, %
one finds \cite{Ref:ScullyLaserQuantumEntropy} %
\begin{align} %
S = k_\text{B} \ln W + \frac{k_\text{B}}{2}, %
\end{align} %
where $W = \sqrt{ 2 \pi ( \Delta n_0 )^2} = \sqrt{2\pi \mathcal{H}}$. %
Note that for $T \rightarrow 0$, we need to use the expression \eqref{Eq:DMBECGnd} %
or the full canonical ensemble calculation. %


\begin{figure}[b] %
\centering %
\includegraphics[width=3.2in]{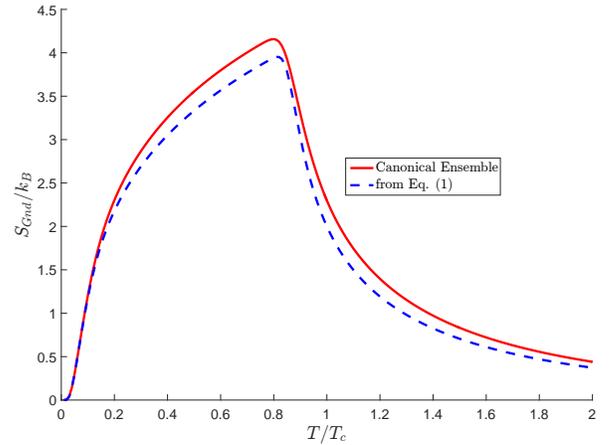} %
\caption{Entropy of Ground state in an Ideal Bose Gas, which is trapped in 3D Harmonic trap. %
The total number of particles is $ N = 200$. %
The critical temperature for 3D harmonic trap is $T_c = \hbar \Omega /k_\text{B} (N/\zeta(3))^{1/3}$, %
with harmonic trap oscillation frequency $\Omega$, and Riemann's zeta function $\zeta(s)$. %
This exact result on entropy is calculated by Canonical Ensemble Partition function, which is explained %
in the Supplementary Information (SI), %
and it is drawn as a solid red line. %
From the approximate density matrix, Eq.~\eqref{Eq:DMBECGnd}, the corresponding von Neumann's entropy is plotted as a dashed blue line. %
} %
\label{Fig:GndEntropy} %
\end{figure} %


In this letter we study the Bose gas in a three dimensional trap. %
We use the canonical ensemble to obtain exact results for the quantum statistical entropy. %
Our exact results reveal new features of the Bose gas. %
We consider the density matrix associated with the ground state $\rho_\text{gnd}$ and for the excited states $\rho_\text{ex}$ obtained from the full canonical density matrix. %
The considerations of exact canonical ensemble reveal that the total entropy of the Bose gas %
at any temperature $T$ is equal to the entropy of the particles in the excited states; %
although the entropy of the ground state particles is nonzero. %
This remarkable result implies the existence of the correlation entropy in Bose gas %
and in fact the correlation entropy must cancel the contribution from the ground state. %
We trace this result to the fact that in the ensemble the number of particles is fixed and thus the total density matrix does not factorize $\rho_\text{T} \neq \rho_\text{gnd} \otimes \rho_\text{ex}$. %
The nonfactorized nature of the full density matrix is further clarified by calculating the correlation functions %
between the ground state and excited state particles. %


\begin{widetext} %


\begin{table}[th]
\centering %
\includegraphics[width=0.7\textwidth]{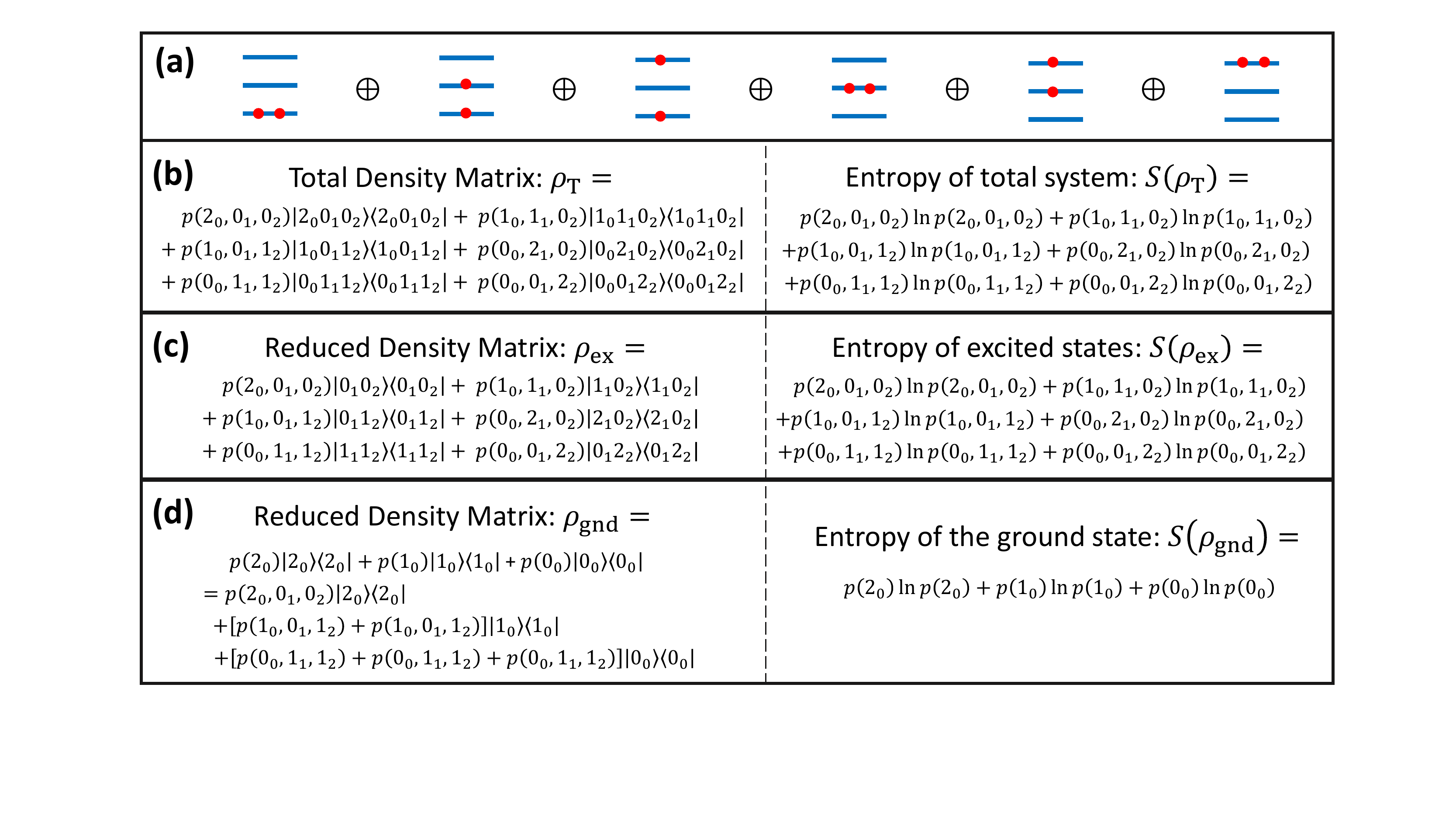} %
\caption{(a) The system consists of two identical Bose particles (red dot), which are distributed among three different states %
(blue lines). %
Due to the Bose statistics, the number of possible configurations is 6. %
(b) The total density matrix $\rho_\text{T}$ and the corresponding entropy $S(\rho_\text{T})$ and
(c) the reduced density matrix $\rho_\text{ex}$ for excited states and the corresponding entropy $S(\rho_\text{ex})$. %
By comparing insets (b) and (c), we can easily confirm the equality of the two entropies. %
(d) The density matrix and entropy for the ground state. %
The relation between the occupation probability for the ground state and the whole joint probability is explicitly shown. %
} %
\label{Table:DiagramDM} %
\end{table} %


\end{widetext} %

\textit{BEC Joint Ground State Entropy}---We first prove that the total entropy of the ideal Bose Gas at a temperature $T$ is %
same as the entropy of excited states of that system. %
At equilibrium, %
the total density matrix for an Ideal Bose gas with \textit{fixed} total number of particles $N$ is given by %
\begin{align} %
\label{Eq:DMBoseTotal}
\rho_\text{T} = & \, \sum_{n_0, \{ n_i \}} p( n_0, \{ n_i \}) %
\vert n_0, \{ n_i \} \rangle \langle n_0, \{ n_i \} \vert %
\, \delta_{N - n_0, \sum n_i}, %
\end{align} %
with the occupation distribution $\{ n_i \}$ on the excited states %
constrained by the condition $\sum_i n_i = N - n_0$. %
The reduced density matrixes for the ground state and for the excited states are %
\begin{subequations} %
\begin{align} %
\label{Eq:DMBoseGnd} %
\rho_\text{gnd} = & \, \text{Tr}_{ \{ n_i \}_{n_0} } ( \rho_\text{T} ) \\ %
= & \, \sum_{n_0} p(n_0) \vert n_0 \rangle \langle n_0 \vert, %
\end{align} %
\end{subequations} %
and %
\begin{subequations} %
\begin{align} %
\label{Eq:DMBoseEx} %
\rho_\text{ex} = & \, \text{Tr}_{n_0} (\rho_\text{T}) \\ %
= & \, \sum_{\{ n_i \}} p \big( n_0 = \sum_i n_i, \{ n_i \} \big) %
\, \vert \{ n_i \} \rangle \langle \{ n_i \} \vert. %
\end{align} %
\end{subequations} %
The occupation probability for the ground state is
\begin{align} %
\label{Eq:ProbGndOcc} %
p(n_0) = \sum_{ \{n_i\} } p(n_0, \{ n_i \} ). %
\end{align} %
Note that the probabilities for the states $\vert \{ n_i \} \rangle$ in $\rho_\text{ex}$ are %
the same joint probabilities as for the states $\vert n_0, \{ n_i \} \rangle$ in $\rho_\text{T}$. %
The explicit example for calculating the corresponding probability is explained %
in Table \ref{Table:DiagramDM}. %

From the von Neumann entropy, Eq.~\eqref{Eq:vonNeumannEntropy}, %
the corresponding entropies are %
\begin{subequations} %
\begin{align} %
S(\rho_\text{T}) = & \, - k_\text{B} \text{Tr}_{n_0, \{ n_i \} } ( \rho \ln \rho) \\ %
\label{Eq:EntropyWhole} %
= & \, - k_\text{B} \sum_{n_0, \{ n_i \}} p( n_0, \{ n_i \} ) \ln  p( n_0, \{ n_i \} ), %
\end{align} %
\end{subequations} %
and
\begin{subequations} %
\begin{align} %
\label{Eq:EntropyExcDef} %
S(\rho_\text{ex}) = & \, - k_\text{B} \text{Tr}_{\{ n_i \} } ( \rho_\text{ex} \ln \rho_\text{ex} ) \\ %
\label{Eq:EntropyExc} %
= & \, - k_\text{B} \sum_{ n_0, \{ n_i \}} p \big( n_0, \{ n_i \} \big) \ln p \big( n_0, \{ n_i \} \big). %
\end{align} %
\end{subequations} %
Showing that the entropy of the total system, Eq.~\eqref{Eq:EntropyWhole}, is equal %
to that for the excited states, Eq.~\eqref{Eq:EntropyExc}, %
since the accessible states and corresponding probabilities are the same. %
Table \ref{Table:DiagramDM} shows this property explicitly for a system of two Bose particles in three non-degenerate levels. %

Similarly, we can write the entropy of ground state. %
\begin{subequations} %
\begin{align} %
\label{Eq:EntropyGndDef} %
S(\rho_\text{gnd}) = & \, - k_\text{B} \text{Tr}_{ n_0 } ( \rho_\text{gnd} \ln \rho_\text{gnd} ) \\ %
\label{Eq:EntropyGnd} %
= & \, - k_\text{B} \sum_{ n_0} p ( n_0) \ln p ( n_0 ). %
\end{align} %
\end{subequations} %

Furthermore, the above result is applicable for any quantum system of identical particles %
including an ideal Fermi atoms in a trap with a \textit{fixed} total number of particles. %
Hence, we can say that the removal of any single state in canonical ensemble preserves the entropy, %
since the total number of particles is fixed by the constraint. %

Since the total entropy of the system is same as that of the excited states, what is learned from this result? %
In a system of $N$ ideal Bose particles, we can divide the system into two parts: %
one is the ground state and the other is the excited states [Eqs.~\eqref{Eq:DMBoseGnd} and \eqref{Eq:DMBoseEx}]. 
It is also possible to define the entropy of each part [Eqs.~\eqref{Eq:EntropyGndDef} and \eqref{Eq:EntropyExcDef}]. %
Since the total density matrix, Eq.~\eqref{Eq:DMBoseTotal}, does \textit{not} factorize %
as, $\rho_\text{T} \neq \rho_\text{ex} \otimes \rho_\text{gnd}$, %
we expect that the entropy of the total system is not the summation of entropy of each part, %
$S(\rho_\text{T}) \neq S(\rho_\text{gnd}) + S(\rho_\text{ex})$, %
and we thus introduce the correlation entropy \cite{Ref:DefCorrelationEntropy} as %
\begin{align} %
\label{Eq:CorrelationEntropy} %
S_\text{cor} (\rho_\text{gnd}, \rho_\text{ex}) \equiv S (\rho_\text{gnd}) + S(\rho_\text{ex}) - S(\rho_\text{T}). %
\end{align} %
Remarkably, since $S(\rho_\text{T}) = S(\rho_\text{ex})$, we see that %
\begin{align} %
\label{Eq:CorrelationEntropyCE} %
S_\text{cor} (\rho_\text{gnd}, \rho_\text{ex}) = S (\rho_\text{gnd}). %
\end{align} %
Therefore, the entropy of ground state can be interpreted as the correlation entropy %
between the ground state and excited states. %
According to information theory \cite{Ref:NielsenChuang}, %
the correlation entropy $S_\text{c} ( \rho_\text{gnd}, \rho_\text{ex})$ %
is called as the mutual information. %
Hence, according to the information theory %
we can say that the status of the excited states can provide total information about the ground state. %

\begin{figure}[th] %
\centering %
\includegraphics[width=3.2in]{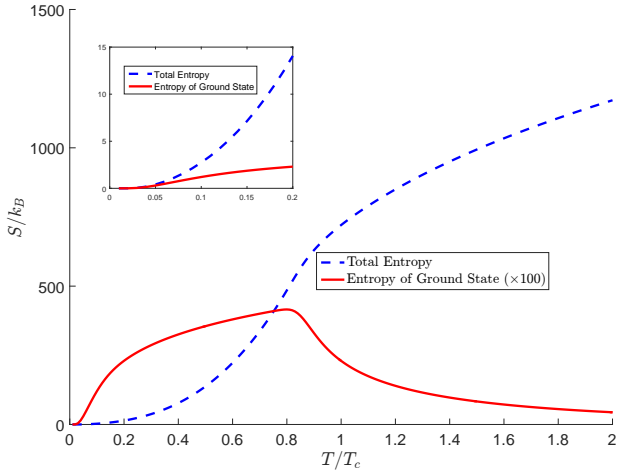} %
\caption{Entropy for ideal Bose Gas which is trapped in a three dimensional harmonic trap. %
The detailed parameters are as in Fig.~\ref{Fig:GndEntropy}. %
The total entropy is drawn with a dashed blue line, using procedure in SI, %
and the entropy of the ground state is in the solid red line. %
In this picture, the entropy for the ground state is multiplied by 100. %
From the behavior of the occupation number in the ground state, %
we can see the entropy contribution of the ground state is important below the critical temperature. %
In a similar way, in terms of correlation entropy the relevant range of the correlation is also below the critical temperature. %
The inset shows that both entropies below $T/T_c = 0.2$. %
} %
\label{Fig:Entropy3DTrap} %
\end{figure} %

\textit{BEC conditional ground state entropy}---In statistics %
and Shannon's information theory \cite{Ref:WildeQuantumInformationTheory}, %
conditional distributions and the conditional entropy are useful concepts. %
Using the conditional probability, we can identify the amount of contribution of the ground state %
in entropy to the excited states. %
The conditional probability for the excited states with a given number of particles in the ground state is %
\begin{align} %
p\big( \{ n_i \} \big \vert n_0 \big) = \frac{ p\big( n_0, \{ n_i \} \big)}{ p(n_0)} %
\end{align} %
where the ground state occupation probability is given by Eq.~\eqref{Eq:ProbGndOcc}. %
The entropy of $\rho_\text{ex}$ can be further evaluated. %
\begin{widetext} %
\begin{align} %
S(\rho_\text{ex}) = & \, - k_\text{B} \sum_{ \{ n_i \}}
\sum_{n_0} \left[ p( n_0)  p\big( \{ n_i \}_{n_0} \big\vert n_0 \big) \right] \log \left[ p( n_0) \right] %
- k_\text{B} \sum_{ \{ n_i \}}
\sum_{n_0} \left[ p( n_0)  p\big( \{ n_i \}_{n_0} \big\vert n_0 \big) \right] \log \left[ p\big( \{ n_i \}_{n_0} \big\vert n_0 \big) \right] \\ %
= & \, - k_\text{B} \sum_{n_0} p(n_0) \log p(n_0) %
- k_\text{B} \sum_{n_0} p(n_0)
\sum_{ \{ n_i \} } p\big( \{ n_i \}_{n_0} \big\vert n_0 \big) \log p\big( \{ n_i \}_{n_0} \big\vert n_0 \big) \\ %
= & \, S(\rho_\text{gnd}) + \sum_{n_0} p(n_0) S ( \rho_\text{ex}^{N-n_0} ). %
\end{align} %
\end{widetext} %
where $\rho_\text{ex}^{N-n_0}$ is the reduced density matrix of excited states with $N - n_0$ particles, %
and $S ( \rho_\text{ex}^{N-n_0} )$ is the corresponding entropy. %
Hence, the excited states $S(\rho_\text{ex})$ contain information about the ground state. %

Similarly, we can rewrite the above relation for the total entropy as %
\begin{align} %
\label{Eq:JointEntropyThm} %
S(\rho_\text{T}) = S(\rho_\text{gnd}) + \sum_{n_0} p(n_0) S ( \rho_\text{ex}^{N-n_0} ). %
\end{align} %
This relation is known as Joint entropy theorem \cite{Ref:NielsenChuang, Ref:OhyaQEntropy, Ref:WehrlEntropyRMP}. %
The entropy contribution of ground state is in the total entropy. %
We can interpret that $S(\rho_\text{gnd})$ as the entropy of the ground state and as the correlation entropy. %

The explicit procedure how to calculate the entropy for the $S(\rho_\text{T})$ and $S(\rho_\text{exc}^{N-n_0})$ is explained in SI. %
Fig.~\ref{Fig:Entropy3DTrap} shows the entropy of ground state, %
or the correlation entropy for the Ideal Bose Gas with 200 particles in 3D harmonic trap. %

\textit{Correlation function}---In order to better appreciate the nature of correlations in the Bose gas at low temperatures, %
we examine the variety of correlations of the occupation numbers between the ground state and the excited states. %
The entropy is defined by the distribution of occupation numbers, that is the density matrix, %
and the correlation function is defined by the corresponding random variables, that is the occupation numbers. %
For the ground state distribution the occupation number $n_0$ for ground state is the corresponding variable, %
and for the excited states the occupation number is $\sum_i n_i = N - n_0$. %

As in statistical description of correlation between two random variables, %
we can introduce the correlation between the numbers of particles in the ground state and in excited states as %
\begin{align} %
\nonumber %
C_1(n_0, \sum_i n_i) \equiv & \, %
\frac{\langle n_0 \sum_i n_i \rangle}{ \sqrt{ \langle {n_0}^2 \rangle \, \langle (\sum_i n_i)^2 \rangle}} \\ %
\label{Eq:CorrelationGndExcC1} %
= & \, \frac{\langle n_0 ( N - n_0 ) \rangle}{ \sqrt{ \langle (n_0)^2 \rangle \, \langle ( N - n_0 )^2 \rangle}}. %
\end{align} %
Note that the Schwarz inequality implies that $C_1 \leq 1$. %
We note that over the temeprature range $T/T_c \sim [0.2 - 0.8]$, $C_1 \simeq 1$ implying %
very high degree of correlation. %
Beyond this temperature the correlation starts falling. %
Next we introduce the correlation defined as the fluctuation around the mean %
\begin{align} %
C_2(n_0, \sum_i n_i) \equiv & \, %
\big[ \langle n_0 \rangle \langle \sum_i n_i \rangle - \langle n_0 \sum_i n_i \rangle \big]^{1/2} \\ %
\label{Eq:CorrelationGndExcC2} %
= & \, \sqrt{ \langle (n_0)^2 \rangle - \langle n_0 \rangle^2}. %
\end{align} %
It is interesting that the conservation of total number $N$ of particles makes $C_2$ identical to the $(\text{variance})^{1/2}$ of the ground state number. %
The $C_2$ shows a behavior which has similarities to the behavior of the correlation entropy. %
However, the correlation entropy shows a much slower dependence on $T$. %
This can be understood as the ground state entropy is the mean value of $p(n_0)$ %
and is related in principle to all order of moments of $n_0$. %
If $p(n_0)$ were to be approximated by a Gaussian, then $\ln p(n_0)$ is directly related to $\ln C_2$ %
and because of logarithmic dependence, entropy shows much slower dependence on $T$ than $C_2$. %
In this figure we also show a very interesting character of the statistics of the fluctuations in the ground state: %
the fluctuations in the region close to $T/T_c \ll 1$ are predominantly sub-Poissonian %
as $\Delta n_0 / \sqrt{\langle n_0 \rangle} < 1$. %
The result from the approximate expression, Eq.~\eqref{Eq:DMBECGnd}, are close to the exact result. %

Although the fluctuations of the ground state populations have not been yet studied experimentally, %
this is possible in principle from the snapshots of the images of the distribution of particles in the trap. %
The peak and tail of the snapshots should yield the ground state and the excited state distributions. %
Such images have been used for studying the particle number fluctuations in a trap %
when interparticle interactions are important \cite{Ref:RaizenExp}. %

\begin{figure}[ht] %
\centering %
\includegraphics[width=3.2in]{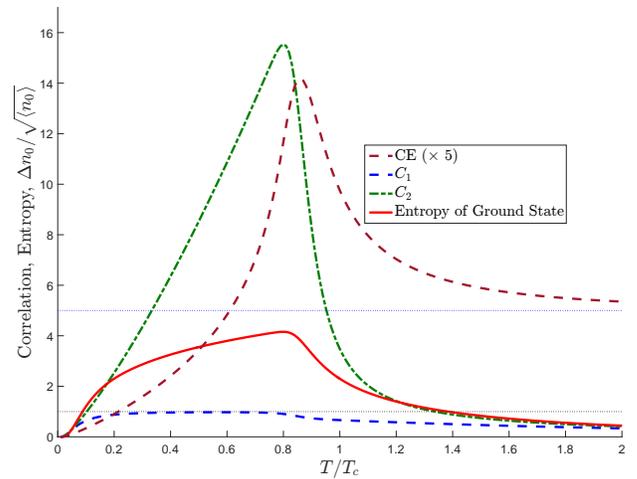} %
\caption{The system is an Ideal Bose Gas trapped in 3D harmonic trap with 200 particles. %
The parameters are the same as in Fig.~\ref{Fig:GndEntropy}. %
The normalized correlation function $C_1$, Eq.~\eqref{Eq:CorrelationGndExcC1}, %
between ground state occupation number and that of excited states is plotted as a dashed blue line. %
$C_2$, Eq.~\eqref{Eq:CorrelationGndExcC2}, is plotted as a dotted green line. %
The correlation entropy, Eq.~\eqref{Eq:CorrelationEntropyCE}, or the entropy of the ground state is also drawn %
as a solid red line. %
In the figure we also show the sub-Poissonian nature of fluctuations %
by plotting the parameter $\Delta n_0 / \sqrt{ \langle n_0 \rangle}$ (dashed brown line [$\times 5$]). %
The strong sub-Poissonian region corresponds to $\Delta n_0 / \sqrt{ \langle n_0 \rangle} \ll 1$. %
} %
\label{Fig:Cor_Gnd_WholeExcited}
\end{figure} %

We next consider the correlation between two specific states defined by %
\begin{align} %
\label{Eq:CorrelationC1} %
\tilde{C}_1(n_i, n_j) \equiv & \, \frac{ \langle n_i n_j \rangle }{ \sqrt{ \langle (n_i)^2 \rangle \langle (n_j)^2 \rangle}} %
\end{align} %
where %
\begin{align} %
\langle n_i n_j \rangle = \sum_{n_i=0}^N \sum_{n_j=0}^{N - n_i} e^{-\beta n_i \epsilon_i - \beta n_j \epsilon_j} \frac{Z_{N - n_i - n_j}(\beta)}{Z_N(\beta)}, %
\end{align} %
which is derived in the supplementary information. %

\begin{figure}[ht] %
\centering %
\includegraphics[width=3.2in]{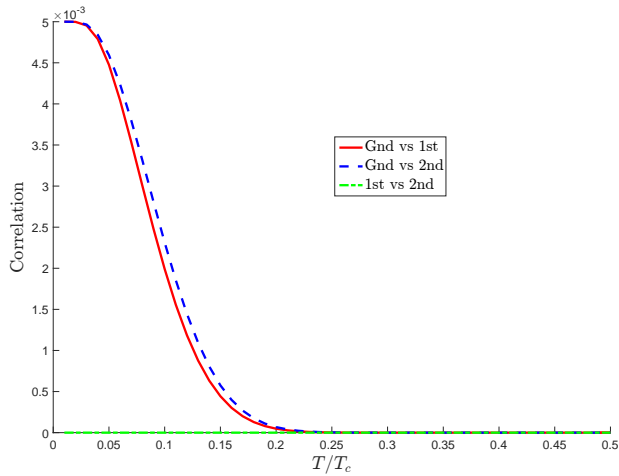} %
\caption{The correlation function $\tilde{C}_1$, Eq.~\eqref{Eq:CorrelationC1}, is drawn among the three lowest states. %
$\tilde{C}_1(n_0,n_1)$ is drawn as a solid red line, $\tilde{C}_1(n_0, n_2)$ as a blue dashed line, and $\tilde{C}_1(n_1,n_2)$ as a green dashed-dotted line. %
Since the occupation of the ground state is macroscopic in low temperature, %
the correlation function is noticeable below $T/T_c \, \sim 0.2$. %
The correlation between the first and the second excited states is negligible, %
since the occupation number in each state is small compared to the total number of particles. %
The system is an ideal Bose Gas trapped in 3D Harmonic trap, %
and the parameters are the same as in Fig.~\ref{Fig:GndEntropy}. %
} %
\label{Fig:Cor_Gnd_1stExcited} %
\end{figure}

The correlation between the ground state and the first excited state is shown in Fig.~\ref{Fig:Cor_Gnd_1stExcited}. %
Though the occupation number of the first excited states is considerable around $T/T_c \sim 1$, %
the correlation between two states are negligible except at low temperatures $T/T_c \lesssim 0.1$, %
where it is of order $1/N$. %

\textit{Summary}---The most important result of our exact calculation based on the canonical ensemble is %
that the entropy of a Bose gas confined to a three dimensional harmonic trap is equal %
to the entropy associated with the atoms in the excited states. %
This is so even though, at any temperature, the entropy of the particles in the ground state is nonzero. %
We bring out the reasons for this surprising result by showing %
that the total entropy associated with the full system consists of three contributions---the entropy of the ground state, %
the entropy associated with the particles in the excited state and  a contribution %
which we refer to as the correlation entropy [analog of the mutual information from information theory]. %
We show on a very general ground that the correlation entropy cancels the ground state contribution. %
This appears due to the fixed number of particles distributed among the quantum states \cite{Ref:CommentFermion}. %
The explicit nature of correlations among the particles in ground state and excited states is brought about %
by studying different types of correlation functions involving the numbers in ground state and excited states. %
Because of the number conservation, these correlations become related to the ground state fluctuations. %
Since the entropy of the ground state is the mean value of the $\log p(n_0)$, %
the fluctuations of $n_0$ determine the value of the entropy of particles in the ground state.

This paper is supported by the Office of Naval Research (Award No. N00014-16-1-3054) and by the Robert A. Welch Foundation (Grant No. A-1261). %


\begin{thebibliography}{99}
\bibitem{Ref:CommentFermion} %
This is also applicable to particles obeying Fermi-Dirac statistics. %
Consequences of this would be the subject of future study. %
\end{thebibliography}
\end{document}


\begin{center} %
\Large %
\textbf{Supplementary Information} %
\end{center} %



\section{\textit{Exact} Partition Function and Occupation Probability in Canonical Ensemble} %
\label{App:ZnPnCE} %

The partition function $Z_N$ in canonical ensemble (CE) can be written in terms of occupation number in each accessible states as %
\begin{align} %
Z_N (\beta) = & \, \sum_{n_0=0}^\infty \sum_{n_1=0}^\infty \cdots \sum_{n_\nu = 0}^\infty \cdots %
e^{-\beta n_0 \epsilon_0} e^{-\beta n_1 \epsilon_1} \cdots e^{-\beta n_\nu \epsilon_\nu} \cdots %
\delta \left( N - \sum_\nu n_\nu \right) \\ %
= & \, \sum_{ n_0, \{ n_i \}_{n_0} } %
e^{- \beta \sum_\nu n_\nu \epsilon_\nu} \delta \left( N - \sum_\nu n_\nu \right), %
\end{align} %
where $\beta = (k_B T)^{-1}$ is the inverse temperature with the Boltzmann constant $k_\text{B}$. %

Let's consider the probability that $\nu$-state has more than $n$ particles. %
Then, the corresponding summation is restricted to $n_\nu \geq n$. %
\begin{align} %
P (n_\nu \geq n) = & \, \frac{1}{Z_N} \sum_{n_0=0}^\infty \sum_{n_1=0}^\infty \cdots \sum_{n_\nu = n}^\infty \cdots %
e^{-\beta n_0 \epsilon_0} e^{-\beta n_1 \epsilon_1} \cdots e^{-\beta n_\nu \epsilon_\nu} \cdots %
\delta \left( N - \sum_\nu n_\nu \right) \\ %
= & \, e^{-\beta n \epsilon_\nu} \frac{Z_{N-n}(\beta)}{Z_N(\beta)}
\end{align} %
The probability for $\nu$ state to have $n$ particles is %
\begin{align} %
\label{Eq:ProbOccupationSingle} %
P(n_\nu = n) = & \, P(n_\nu \geq n) - P(n_\nu \geq n+1) \\ %
= & \, \frac{ e^{-\beta n \epsilon_\nu} Z_{N-n}(\beta) - e^{-\beta (n+1) \epsilon_\nu} Z_{N-n-1}(\beta)}{Z_N(\beta)}. %
\end{align} %
The average occupation number in the state $\nu$ is %
\begin{align} %
\langle n_\nu \rangle = & \, \sum_{n_\nu = 1}^N n_\nu P(n_\nu) %
= \sum_{n_\nu = 1}^N e^{- \beta n_\nu \epsilon_\nu} \frac{Z_{N-n_\nu}(\beta)}{Z_N(\beta)}. %
\end{align} %
And, the total number of particles is given by sum of the average occupation number of all states %
\begin{align} %
N = & \, \sum_\nu \langle n_\nu \rangle. %
\end{align} %
By a simple manipulation, we will get the following recurrence relation \cite{Ref:LansbergThermodynamics, Ref:WeissWilkensOE1997}. %
\begin{align} %
Z_N(\beta) = & \, \frac{1}{N} \sum_{m=1}^N Z_1(m \beta) Z_{N-m} (\beta). %
\end{align} %

Similar to Eq.~\eqref{Eq:ProbOccupationSingle}, we can write the occupation probability for two states %
\begin{align} %
P( n_\nu \geq n, n_\mu \geq m) = & \, e^{-\beta n \epsilon_\nu - \beta m \epsilon_\mu} \frac{Z_{N-n-m}(\beta)}{Z_N(\beta)} %
\end{align} %
So, the probability to find $n_\nu = n$ and $n_\mu = m$ is %
\begin{align} %
& \, P( n_\nu = n, n_\mu = m) \\ %
= & \, P(n_\nu \geq n, n_\mu \geq m) - P(n_\nu \geq n, n_\mu \geq m+1) \\ %
& \, \quad - P(n_\nu \geq n+1, n_\mu \geq m) + P(n_\nu \geq n+1, n_\mu \geq m+1) %
\end{align} %
The correlation function between the two states can be easily obtained. %
Explicitly, it is %
\begin{align} %
\label{Eq:CorrelationTwoStates} %
\langle n_\nu n_\mu \rangle %
= \sum_{n=1}^N \sum_{m=1}^N e^{-\beta n \epsilon_\nu - \beta m \epsilon_\mu} \frac{ Z_{N-n-m}(\beta)}{Z_N(\beta)}. %
\end{align} %

\section{Thermodynamic Quantities in Canonical Ensemble} %
\label{App:ThermodynamicQuantitiesCE} %

Partition function $Z_N(T,V)$ in canonical ensemble is related to the Helmholtz free energy $A(T,V)$ %
\cite{Ref:Pathria3rd, Ref:Huang2nd} %
\begin{align} %
Z_N(T,V) = & \, e^{-\beta A(T,V)} \\ %
\intertext{or} %
A(T,V) = & \, - k_B T \ln Z_N(T,V) %
\end{align} %

Thermodynamic quantities can be calculated from the Helmholtz free energy through the Maxwell relations. %
For example, pressure $P$ and entropy $S$ are %
\begin{align} %
P = & \, - \left( \frac{\partial A}{\partial V} \right)_T \\ %
S = & \, - \left( \frac{\partial A}{\partial T} \right)_V \\ %
U = & \, \langle H \rangle = A + TS \\ %
C_V = & \, \left( \frac{\partial U}{\partial T} \right)_V %
\end{align} %
with isochoric heat capacity $C_V$. %

In terms of partition function, %
\begin{align} %
\frac{A}{k_B} = & \, - T \ln Z_N \\ %
\frac{P}{k_B T} = & \, \left( \frac{\partial \ln Z_N}{\partial V} \right)_T \\ %
S = & \, k_B \ln Z_N + k_B T \left( \frac{\partial \ln Z_N}{\partial T} \right)_V \\ %
U = & \, k_B T^2 \left( \frac{\partial \ln Z_N}{\partial T} \right)_V \\ %
C_V = & \, 2 k_B T \left( \frac{\partial \ln Z_N}{\partial T} \right)_V %
+ k_B T^2 \left( \frac{\partial^2 \ln Z_N}{\partial T^2} \right)_V %
\end{align} %
Derivatives of the partition function $\ln Z_N$ with respect to the temperature $T$ or to the volume $V$ give %
to the corresponding thermodynamic quantities in canonical ensemble. %